\documentclass[prd,tightenlines,twocolumn]{revtex4}
\usepackage{amsmath}
\usepackage{amssymb}
\usepackage{graphicx}
\usepackage{bm}
\usepackage{enumerate}
\usepackage{color}
\newcommand{\be}{\begin{equation}}
\newcommand{\ee}{\end{equation}}
\newcommand{\bea}{\begin{eqnarray}}
\newcommand{\eea}{\end{eqnarray}}
\newcommand{\ba}{\begin{eqnarray}}
\newcommand{\ea}{\end{eqnarray}}

\begin{document}

\title{ Regimes of the Pomeron and its Intrinsic Entropy }
\author{ 
 Edward Shuryak and Ismail Zahed}

\affiliation{Department of Physics and Astronomy, \\ Stony Brook University,\\
Stony Brook, NY 11794, USA}


\begin{abstract} 
We suggest that the perturbative and non-perturbative descriptions of the Pomeron 
can be viewed as complementary descriptions of different phases in the Pomeron 
phase diagram, with a phase boundary where the proper description of
the produced systems are  ``string balls". Their
intrinsic entropy  is calculated and turned out to be the same, as the recently reported perturbative entanglement
entropy.  The distribution of large multiplicities stemming from the string balls is
also  wide, with its
moments  close to those reported for hadrons in $pp$ collisions at the LHC. 
At low-x, the quantum string is so entangled that sufficiently weak string 
self-attraction can cause it to turn to a string ball dual to a black hole. We suggest that
low-x saturation occurs when the density of wee-strings reaches the Bekenstein bound, 
with a proton size that freezes with increasing rapidity.  Some of these observations maybe  checked
at the future eIC.
 \end{abstract}

\maketitle

\section{Introduction}

Already in the 1960's high energy hadronic collisions were described
using ``Reggeon exchanges" with various quantum numbers. 
The $Pomeron$, named after Pomeranchuck who introduced  the leading exchange with vacuum quantum numbers,
 dominates
 hadronic collisions at  high energies.  Phenomenological descriptions of weakly interacting Pomerons
 have been developed by Gribov and collaborators, see  \cite{Gribov:1973jg,Donnachie:1992ny}. 

In the 1970's, with the advent of QCD in its weak coupling form,  
a lot of work has been devoted to describe high energy collisions by re-summing
certain gluonic Feynman diagrams. This program has been, to leading order, 
completed by Balitsky, Fadin, Kuraev and Lipatov \cite{Kuraev:1977fs}
and is known as the BFKL Pomeron. Reformulation of it in terms of Wilson loops, and the addition of the nonlinar effects
leading to saturation, has lead to the so called BK equation, due to  Balitsky and Kovchegov \cite{Balitsky:1995ub}. 

While the perturbative description is valid at small distances, hadronic collisions 
deal with object sizes and impact parameters $\sim 1\, {\rm fm}$, where nonperturbative effects due to confinement are dominant. Therefore multiple efforts have been made to 
develop a ``non-perturbative Pomeron". 
In this work we discuss one of such approaches, developed using semiclassical
tunneling and an effective long string action by Basar, Kharzeev, Yee and Zahed \cite{Basar:2012jb},
for brevity to be called BKYZ Pomeron. Its main elements will be presented in the next section.

(We will not review other versions of  non-perturbative Pomerons, and just note in passing
that a holographic idea relating the Pomeron to a dual graviton exchange
\cite{Brower:2006ea} has evolved into a rather successful theory  
of double-diffractive production \cite{Domokos:2009hm,Iatrakis:2016rvj} in a framework of 
AdS/QCD.)


In a previous paper by the two of us~\cite{Shuryak:2013sra}, to be referred to  below as I, it was pointed out that the stringy Pomeron possesses an {\em intrinsic temperature and entropy}. It happens
because the classical world-volume of the exchanged string, for brevity to be called a``tube", 
possesses a periodic coordinate, which can be identified with a Matsubara time. Therefore
quantum oscillations of the tube have the form of a thermal theory.
This temperature
depends on the location along the tube, its maximal value is
\be 
{1 \over T}=\beta= {2\pi b \over \chi}, \label{eqn_beta} 
\ee 
where $b$ is the impact parameter (the length of the tube) and $\chi={\rm ln}(s/s_0)$
is the relative rapidity of the beams.
In the standard way, this temperature defines  the energy, entropy and  other
thermodynamic quantities of the system.

Further arguments in I  point  out that 
since the QCD strings are well known to exhibit
the so called Hagedorn transition
as a function of  temperature,
real or ``effective",  at a certain temperature $T_H$. As $T\rightarrow T_{H}$
from below, the string gets excited and becomes very long. Its energy $E$
is however cancelled by the entropy term in the free energy $F=E-T S$, keeping $F$ small. Such behavior of strings as a function of $T$ in pure gauge theories has been studied in detail, with the conclusion that $T_H$ is only slightly higher than the critical temperature
$T_c\approx 270\, {\rm MeV}$ (for the $SU(3)$ color group). Below we will find that the Pomeron as a twisted tube carries
a lower intrinsic Hagedorn temperature. 

Studies of the so called ``elastic scattering profile function"  $F(b)$ in I,  have identified
three distinct regimes: (i) a ``cold string" at large $b$, in which $F(b)\sim {\rm exp}(-b^2)$ 
due to the dominant classical action of the string; (ii) a ``near-critical string" at intermediate $b$,
in which the amplitude grows as $F(b)\sim {\rm exp}(-\sqrt{1-b^2/b_c^2})$;
and an ``over-excited string" at $b<b_c$ or $T>T_H$ in which the nucleon is effectively black 
with $F(b)\sim 1$. All three regimes are clearly seen in the LHC data on $F(b)$, 
see Fig. 4 of I.

While the paper I was devoted to description of the $elastic$ $pp$ collisions, we now try to extend the
notion of three regimes of the Pomeron to the $inelastic$ collisions. But, before
we proceed with this task, we need to mention several important works which influenced our thinking.

The highly excited string state has been described in~\cite{Kalaydzhyan:2014tfa,Qian:2015boa}
in terms of the so called {\em string balls}. This notion which  originated from the string theory literature,
describes a self-interacting string system that interpolates between a free  string
at small mass, and a black hole at large mass. We will continue along this line in section VI.C
below. 

Another development,  triggering the present paper, is due to Kharzeev and Levin
\cite{Kharzeev:2017qzs}. In this work, devoted to a perturbative BK-like description of the Pomeron, 
the authors pointed out that the produced system of gluons has certain
distribution over the gluon number $P_N$ and thus certain {\em intrinsic entropy}
(but no temperature!). Let us for clarity explain that the term ``intrinsic" here and elsewhere
is used to emphasize that it is developed prior to the collision, to distinguished it from the 
``final entropy" related to the system of hadrons observed in the detectors. 
 While $P_N$ itself has been derived previously, Kharzeev and Levin  successfully 
compared it to the distribution over $hadronic$ multiplicity in $pp$ collisions. They have also noted that since 
the gluon production is modeled by a kinetic equation, one can make statements about
the entropy growth as a function of the effective evolution time, or ${\rm ln}(s)$, leading 
eventually to a state of maximal possible entropy. 

The most  important feature  is that 
 the two versions of the Pomeron, BFKL and BKYZ ones, starting from very different
 Lagrangians and
views on the underlying dynamics, end up with the very same expressions for the Pomeron elastic amplitude (modulo parameters). 
The aim of this paper is to discuss the similarities and differences between these two theories,
extending the discussion to $inelastic$ collisions. The central stage in it will be 
taken by the notion of the ``intrinsic entropy", the notion of ``string bits" and the distributions
over them.

\section{The BKYZ Pomeron} 

\begin{figure}[htbp]
\begin{center}
\includegraphics[width=6cm]{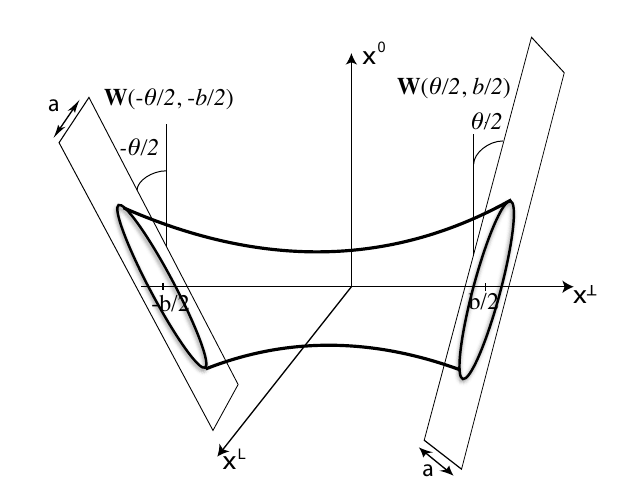}
\includegraphics[width=6cm]{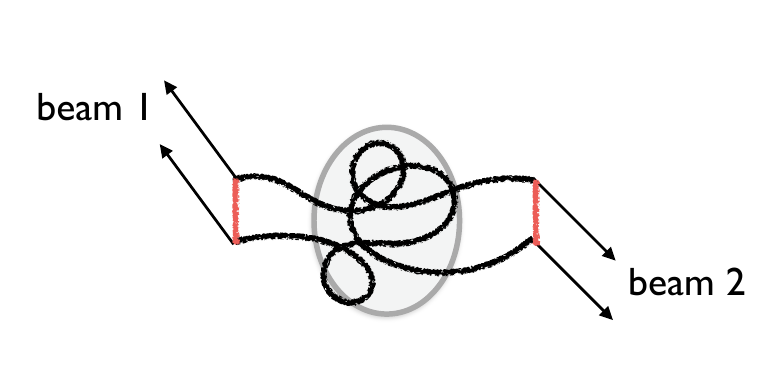}
\caption{The BKYZ Pomeron as a close quantum string exchange between two dipole sheets in Euclidean space (top).
The same exchange when cut horizontally (bottom) from~\cite{Shuryak:2013sra}. }
\label{fig_tube}
\end{center}
\end{figure}

The perturbative BFKL Pomeron is based on a partonic shower, with the
elementary process being a splitting of a gluon into two gluons. The formulation of Mueller
and his collaborators~\cite{Mueller:1994jq} modifies it into a splitting of one dipole into two,
by production of pair of charges.  
The ``stringy" BKYZ Pomeron describes the same basic process of a charge  pair creation, 
but in the confining phase of QCD, in which most of their energy is in the confining strings (the QCD flux tubes).

Let us sketch the main steps of the derivation of the BKYZ Pomeron amplitude
in a graphical form. The upper plot in Fig.\ref{fig_tube}, from \cite{Basar:2012jb}, 
is a sketch of the string world volume describing its  Euclidean time history,  for brevity to be called a ``tube". 
It is a minimal area surface (think of a twisted soap film) interpolating between the
two ``holes" produced in the rectangular world-volume spanned by the
strings, residing in both passing dipoles. 
Cut vertically, it is understood as a virtual exchange of a closed string (a virtual glueball)
between the dipoles. Cut horizontally (see the lower plot in Fig.\ref{fig_tube}, also from \cite{Basar:2012jb}) 
it is a process creating  a pair of strings. 

The derivation starts with the standard approximation, in which the protons are viewed
as  quark-diquark dipoles, ``frozen" at high energies and  represented by Wilson loops
running along two respective light cones. Specifically, the scattering
amplitude for the process $1+2\rightarrow 3+4$ at large $\sqrt{s}$ factorizes~\cite{Nachtmann:1991ua}

\begin{eqnarray}
\label{XDD1}
&&{\mathcal T}_{12\rightarrow 34}(s,t) =-2is \int \frac {dz_1}{z_1}\frac {dz_2}{z_2}  \nonumber\\
&&\times \psi_{34}(z_1)\psi_{12}(z_2) {\mathcal T}_{DD}(\chi,{\bf q}_\perp,z_1,z_2) 
\end{eqnarray}
where $z_{i}$ is related to the transverse size of the i-dipole element described by the wave function $\psi_i$.
The dipole-dipole scattering amplitude is given by

\begin{eqnarray}
\label{XDD2}
{\cal T}_{DD}(\chi,{\bf q}_\perp,z_1,z_2) 
= \int d{\bf b}_\perp \ e^{i {\bf q}_\perp \cdot {\bf b}_\perp} \, {\bf WW} 
 \label{dipdip6}
\end{eqnarray}
with the Wilson loop correlator

\be
\label{XDD3}
{\bf WW}\equiv \langle {\bf W}(C_1) {\bf W}(C_2) \rangle -1
\ee
and  the normalization $\langle {\bf W}\rangle=1$.
The Wilson loops are evaluated along closed rectangular surfaces $C_{1,2}$ lying on the light cone,
at an impact parameter $b$. Fig.~\ref{dipdip6} gives an illustration of the set up in Euclidean space with 
the identification $\theta\rightarrow i\chi$ in Minkowski space. 
The averaging in (\ref{XDD3}) is over the Yang-Mills gauge fields.  
This correlator can be calculated either in 4-d flat space, or, using holography at strong coupling, in 5-d deformed $AdS_5$.

For 2 incoming dipoles of identical size $a_D$  the  result for $b>\beta$ is~ \cite{Basar:2012jb,Zahed:2012sg}

\be
{\bf WW}\equiv -\frac{g_s^2a_D^2}{4\alpha^\prime}\,{\bf K}_T(\beta, b)
\label{cor1}
\ee
where the transverse partition function

\begin{eqnarray}
\label{14X}
&&{\bf K}_T\left(\beta, b\right)=\nonumber\\&&\left(\frac{\pi}{\chi}\right)^{\frac{D_\perp}2}
e^{-\sigma\beta b}\,{\rm Tr}\left(e^{-2\chi\,\left({\bf L}_0-\frac{D_\perp}{24}\right)}\right)
\end{eqnarray}
sums up the transverse oscillator modes in flat $D_\perp$, and $\sigma_T=2\sigma$. Here $g_s$ is the string coupling.
The tracing is over the eigenmodes of the tube, over the zero point oscillations
(known as the tachyon) plus the transverse excitations. 
The normal ordered  Virasoro generator  ${\bf L}_0$  
\be
{\bf L}_0=\sum_{n=1}^\infty\sum_{i=1}^{D_\perp}\, : a_{-n}^ia_n^i :
\label{14}
\ee
plays the role of the Hamiltonian, with $a^i_n$ satisfying the  transverse oscillator algebra
\be
\left[a_n^i, a_m^j\right]= n \,\delta^{ij}\delta_{n+m,0}
\label{OSCI}
\ee
The normal ordering in (\ref{14}) produces the zero-point contribution with the
central charge of the bosonic string $C=D_\perp$, the number of transverse dimensions
(2 or 3, depending on the setting). The 
excited modes correspond to  the so called Pomeron  daughters (see a  brief discussion of
those in Appendix B).

   \begin{figure}[t!]
\begin{center}
\includegraphics[width=8cm]{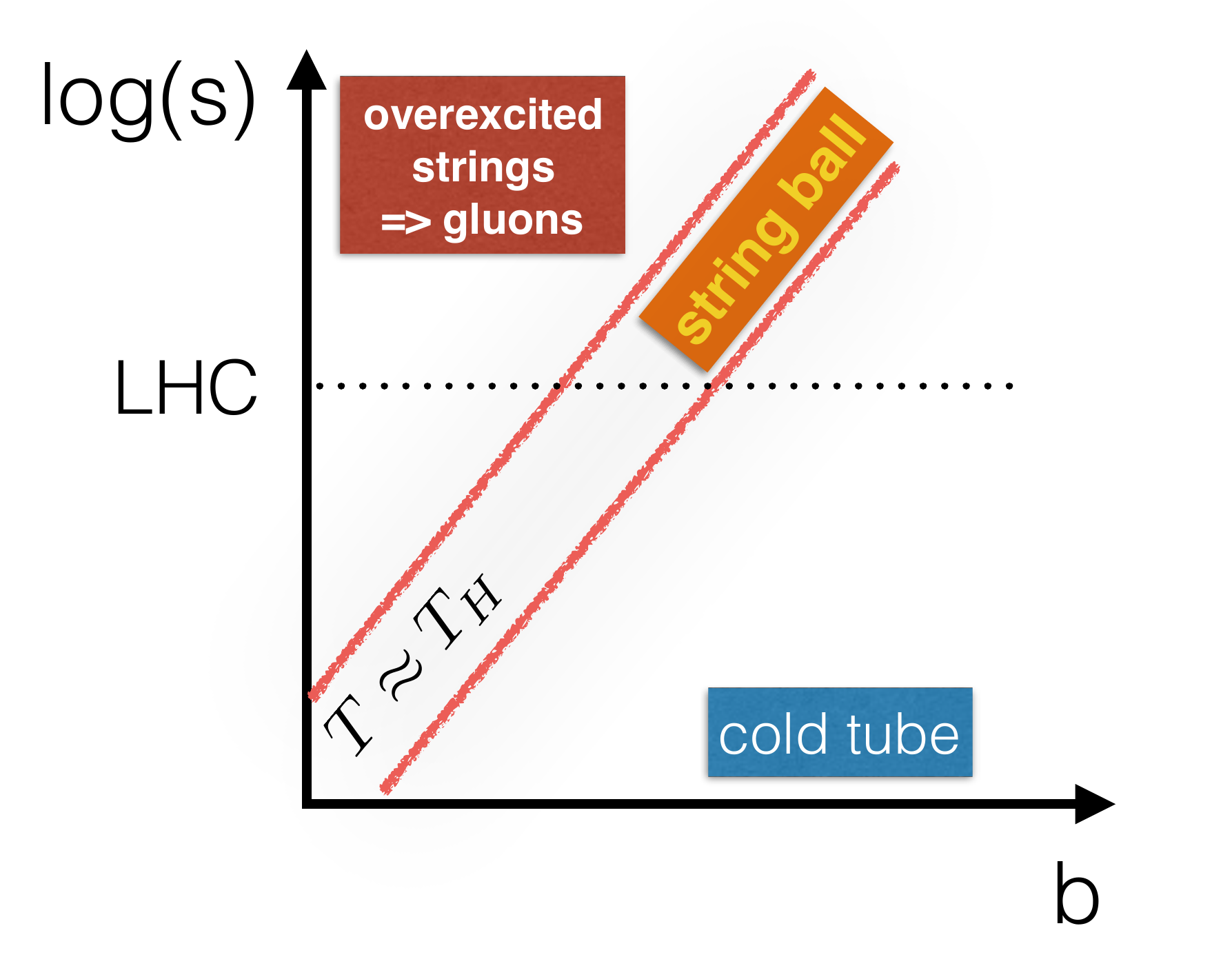}
\caption{The Pomeron ``phase diagram".}
\label{fig_b-chi_plot}
\end{center}
\end{figure}

In pQCD the physical picture explaining why  the hadronic cross sections grow with energy, is known as Gribov diffusion. In it $\chi={\rm ln}(s/s_0)$   plays the role of time: as it grows, the gluons spread in the transverse plane diffusively $r_\perp^2\sim \chi$.
The stringy Pomeron predicts   the same phenomenon, and in fact  the amplitude (\ref{14X})  satisfies the   diffusion equation
\be
\left(\partial_\chi-\Delta-{\bf D}\nabla_{\bf b_\perp}^2\right){\bf K}_T=0
\label{14X1}
\ee
with the Pomeron intercept $\Delta=\alpha_P-1$
\be
\Delta=\frac {D_\perp}{12}-D_\perp\sum_{n=1}^\infty \frac{n}{e^{{2\chi n}}-1}\rightarrow \frac {D_\perp}{12}
\label{14X2}
\ee
and a diffusion constant ${\bf D}=\alpha^{\prime}/2$. (Note that $\alpha'=0$ in pQCD.)

Furthermore, a holographic approach promotes this diffusion, from the two-dimensional
transverse plane to a three-dimensional space $D_\perp=3$, including the holographic coordinate $z$. If 
  appropriately modified by dilaton background, it   can nicely describe the transition from 
the weak coupling  at small $z$, to strong coupling and confinement at large $z$~\cite{Gursoy:2010fj}. A 
consequence is the following correction to
 the Pomeron intercept
\be
\label{INTERCEPT}
\Delta\rightarrow \Delta(\lambda)=\Delta-\frac{(D_\perp-1)^2}{8\sqrt{\lambda}}
\ee
with large  $^\prime$t Hooft coupling $\lambda=g^2N_c$ in the  denominator.

By approximating it by a conformal AdS space, and identifying the size of the dipoles with
their locations in $z$, Stoffers and Zahed \cite{Stoffers:2012zw} generalized the stringy Pomeron
to the treatment of deep inelastic $ep$ scattering, in which one of the dipoles -- originated from a highly virtual photon with large $Q^2$ -- has a very small size $\sim 1/Q$. They have shown that this generalization can describe well the famous HERA data.

 \section{The  ``phase diagram" of the BKYZ Pomeron  } 
 
 The  effective temperature of the string world-sheet is identified with the  inverse $\beta$
already given above (\ref{eqn_beta}). One physical interpretation
of it can be made by noticing that  the string tension causes its end-points to recede from each other with a relative acceleration
$a_U=\chi/b$, thus inducing the so called Unruh temperature $a_U/2\pi$ associated with accelerated frames on the string world-sheet. 
Note that the effective temperature is larger for a small impact parameter $b$ 
(central collisions) or large relative rapidity $\chi$ (large boosts). 

The expressions above were  established  using the Polyakov-Luscher action for long strings with $b\gg \beta$
 in flat $2+D_\perp$ dimensions, for which the tube fluctuations are small.
 However, as $T$ approaches the  Hagedorn temperature here  defined using $\sigma=\sigma_T/2$
 
\be
\label{HAGEDORN}
T_H=\frac 1{\beta_H}=\left(\frac{3\,\sigma}{\pi D_\perp}\right)^{\frac 12}
\ee
the fluctuations increase strongly. For $D_\perp=3$ and $\alpha^\prime=1/2\pi\sigma_T\sim 1/{\rm GeV}^2$,
the intrinsic Hagedorn temperature is $T_H\sim 160 $ MeV for  the Pomeron, which is lower than the
Yang-Mills Hagedorn temperature $\sim 270$ MeV as we noted earlier. It is remarkaby close to the
reported QCD chiral cross-over transition temperature. In this regime, 
 using the Nambu-Goto action instead, we have derived the re-summed action~\cite{Shuryak:2013sra},

\begin{eqnarray}
\label{14X0}
&&{\bf K}_T\left(\beta, b\right)\rightarrow \nonumber\\
&&\left(\frac{\pi}{\chi}\right)^{\frac{D_\perp}2}
{\rm Tr}\left(e^{-\sigma\beta b\left(1+\frac{24}{D_\perp}\frac{\beta_H^2}{ \beta^2}\left({\bf L}_0-\frac{D_\perp}{24}\right)\right)^{\frac 12}}\right)\nonumber\\
\end{eqnarray}
 Its dual relation to the static potential, especially its Arvis form, is  discussed  in Appendix A.
Note that (\ref{14X0}) returns to (\ref{14X}) in the ``cold" regime with $\beta\gg \beta_H$
and $2b>\beta_H$. 
 
%

  For deep inelastic scattering (DIS) one often uses the kinematic range plot of the ${\rm ln}(Q^2)- {\rm ln}(\frac 1x)$ variables, 
where  the regions of weakly coupled vs strongly coupled domains, and saturated vs 
 dilute parton ensembles are identified.
  Let us summarize
this section   with a similar diagram for  the ``Pomeron phase diagram". It is in quotation marks because  the system in question -- two colliding hadrons or dipoles --
 is not really macroscopic, so there is no true phase transition related to thermodynamical singularities
 in this phase diagram.
  
  Our plot, shown in Fig.~\ref{fig_b-chi_plot}, 
 is in the $b-{\rm ln}(s/s_0)$ plane.  The transition region in the middle corresponds to the near-critical regime
  defined by the equation $T\approx T_H$ where $T$ was given in (\ref{eqn_beta})
  and the intrinsic Hagedorn temperature for the Pomeron in (\ref{HAGEDORN}).
While at LHC energies, indicated by a horizontal dotted line, all three regions are clearly visible, at lower ones  (RHIC/ISR colliders) only the ``cold" regime is present and 
the elastic collisions profile is basically Gaussian. For orientation, at LHC the gray transition
region is for $b$ between $0.5$ and $1\,{\rm fm}$. This is to be compared with the ``typical
impact parameter $\sqrt{\sigma_{tot}/\pi}\approx 1.5\, {\rm fm}$. Note further, that while all three regions have comparable size in terms of $b$,  their contribution to the cross section 
scales as $b^2$ weighted with the profile, and therefore the central black region still provides
a rather small contribution to it.

   \section{The  intrinsic entropy  of the Pomeron }

 In QED a moving charge or dipole carries   "Weizsaecker-Williams" field with it.
 In perturbative QCD it is substituted by a cloud of gluons, produced by subsequent
 gluon splittings. As pointed out by Kharzeev and Levin \cite{Kharzeev:2017qzs},
 the perturbative gluon cascade  leads to some rather interesting results, which we
 briefly summarize now.
 
First, they found that  the (von Neumann) entropy is just linear in  $\chi={\rm ln}(s/s_0)$, with the coefficient
being the Pomeron intercept 
 \be 
 {\cal S}(x) =\Delta\,\chi\equiv  \Delta\,{\rm ln}\left({1 \over x}\right) 
 \label{eqn_S_KL}
 \ee
Second, the partonic cascade leads to
 a very wide distribution over the gluon number $n$, with a tail toward large multiplcities

\begin{eqnarray}
P_n(\chi)=&&e^{-\Delta \chi}(1 - e^{-\Delta \chi})^{n-1}\nonumber\\
=&&{1 \over \bar n} \left(1-{1\over \bar n} \right)^{n-1}
 \end{eqnarray}
of the form known in statistics as a negative binomial distribution.

The third important point made by Kharzeev and Levin is that this distribution in fact describes 
the distribution over {\em hadron multiplcity} observed in $pp$ collisions at LHC. This statement is demonstrated by calculating the  moments of the distribution
\be C_q={ \left<n^q\right> \over    \left<n\right>^q } \ee
Their experimental values  $$C^{exp}_2=2.0\pm 0.05, \,\,\, C^{exp}_3=5.9\pm 0.6, $$
$$ C^{exp}_4=21\pm 2,  \,\,\,C^{exp}_5=90\pm 19$$ 
are
close to their theoretical predictions      $C_2\approx1.83, C_3\approx5.0, C_4\approx18.2, C_5\approx 83.$ This observation is proposed   as an evidence for the final hadrons being produced from
the gluon cascade.

Let us now proceed to the issue of the intrinsic entropy of the stringy tube. We will give
first the most direct (thermal) derivation of it.
 The thermal free energy  associated with  the stringy tube is 
${\bf F}=-T{\rm ln}{\bf K}_T$, for the  temperature $T=1/\beta$.  The 
corresponding thermal entropy is, by definition,
\be
{\cal S}_P=\beta^2 \frac{\partial}{\partial\beta}\left(-\frac{{\rm ln}{\bf K}_T}{\beta}\right)
\ee
In flat space, we can differentiate  (\ref{14X}) 
and obtain
\begin{eqnarray}
\label{16XX}
{\cal S}_P
&=& D_\perp\sum_{n=1}^\infty{\rm ln}\left(1+\frac 1{e^{2\chi n}-1}\right)\nonumber\\
&+&2\chi\left(\frac{D_\perp}{12}-\sum_{n=1}^\infty \frac n{e^{2\chi n}-1}\right)\nonumber\\
& -&\frac{D_\perp}{2}\left(1+{\rm ln}\left(\frac{2\chi}{2\pi}\right)\right)
\end{eqnarray}
At large collision energy $\chi \rightarrow \infty$ the leading contribution  is the linear term in
 the  second line 
\be
{\cal S}_P\approx  2{\chi}\frac{D_\perp}{12}\rightarrow 2\chi\Delta (\lambda)
\equiv {2\,{\rm ln}{{\bf N}}}
\label{16X}
\ee
(with the rightmost relation following from the diffusion in curved space as will be detailed shortly).

The derived entropy corresponds to the elastic amplitude, with a ``tube" or two exchanged strings. 
The optical theorem tells us that it corresponds to the cross section, or
the inelastic amplitude $squared$. This means that the inelastic amplitude has $one$ string,
or half a Pomeron,
and thus half of the entropy. We thus conclude that the intrinsic entropy of the 
inelastic $pp$ collision is \be {\cal S}_{\rm inelastic} = \chi\Delta (\lambda)\ee
Note, that this
 main result, the  intrinsic entropy 
of the stringy Pomeron,  turns out to be  {\em the same }
as that of the gluon cloud (\ref{eqn_S_KL}), provided that the Pomeron intercept 
$\Delta$ is changed appropriately.

\section{Distribution of string bits}
In the original stringy Pomeron approach only the elastic amplitude has been calculated,
via a semiclassical ``tube" in Euclidean time describing the tunneling event. At this level, the intrinsic temperature and  entropy we discussed above are just technical parameters describing quantum oscillations of the tube. Elastic scattering  still deals with one quantum state, and thus zero entropy. 

In order to proceed to inelastic collisions, one needs 
to develop certain analog of the ``Cutcosky's cutting rules", allowing 
to implement unitarity and representing the imaginary part of the elastic amplitude as 
a total cross section, the sum over the probabilities to produce all physical final states.  
We have already shown it schematically, as a transition from the upper to the lower sketch in Fig.\ref{fig_tube}.

After the intrinsic entropy is derived, 
 let us now propose a distribution over the number and locations of the ``string bits"
 (or ``wee-strings") produced by 
 the BKYZ Pomeron.


\begin{figure}[t]
  \begin{center}
  \includegraphics[width=8cm]{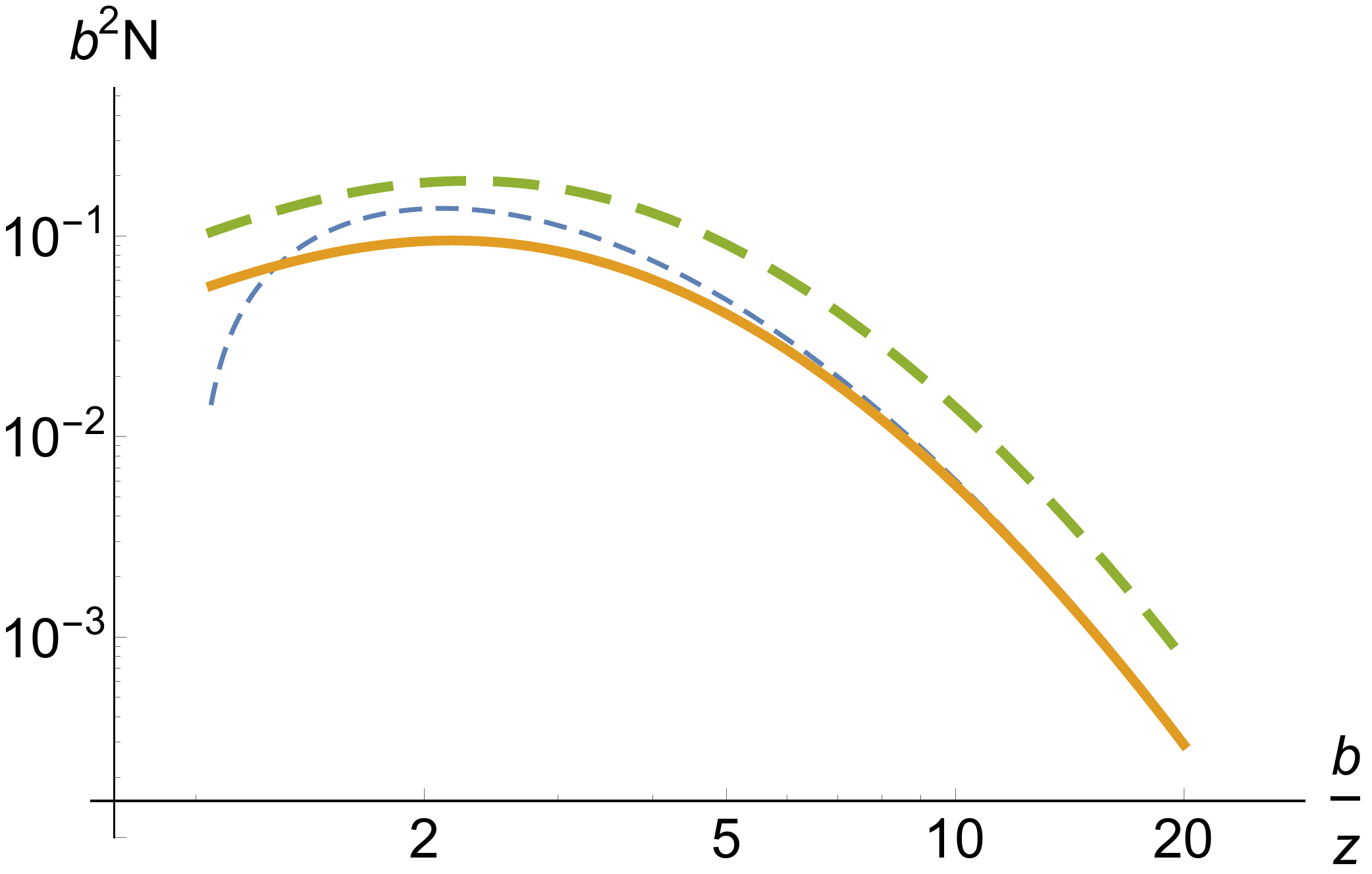}
  \caption{Density distribution  of string bits of fixed size $z=0.9\,z_0$ at a transverse distance $b/z$,
  as sourced by a dipole of size $z^\prime=0.9\,z_0$ at the origin at a relative rapidity 
  $\chi=10$.
  The middle-blue-dashed curve is (\ref{WEEX}) and follows from the  large b-asymptotic of the lower-orange-solid
   curve for AdS without a wall 
  (first contribution in (\ref{NEU})). The  upper-green-long-dashed curve is the full (\ref{NEU}) for AdS with a wall.}
    \label{densitycomp}
  \end{center}
\end{figure}

 We propose to quantify this  in the following way. Keeping in mind
 the holographic extension of the stringy Pomeron already mentioned, one needs to
 ``amputate" the string ends, associated with the colliding dipoles. Technically it leads to 
 the (dimensionless) {\em density of string-bits}
identified with       
\be
{\bf N}\approx \frac{z_0^2{\bf K}_T}{zz^\prime} \label{eqn_N}
\ee
with $0\leq z\leq z_0$ identified as the holographic direction, as we recall
in Appendix C. It satisfies a diffusion equation as well, and therefore also satisfies
the
``chain rule"
\begin{eqnarray}
{\bf N}(3,1)=\int d2\,{\bf N}(3,2)\,{\bf N}(2,1)
\label{HW10}
\end{eqnarray}
It is a necessary feature of a cascade, which can always be 
split into two subsequent cascades if some ``measurement" needs to be done, at some intermediate time.

\begin{figure}[t]
  \begin{center}
  \includegraphics[width=6cm]{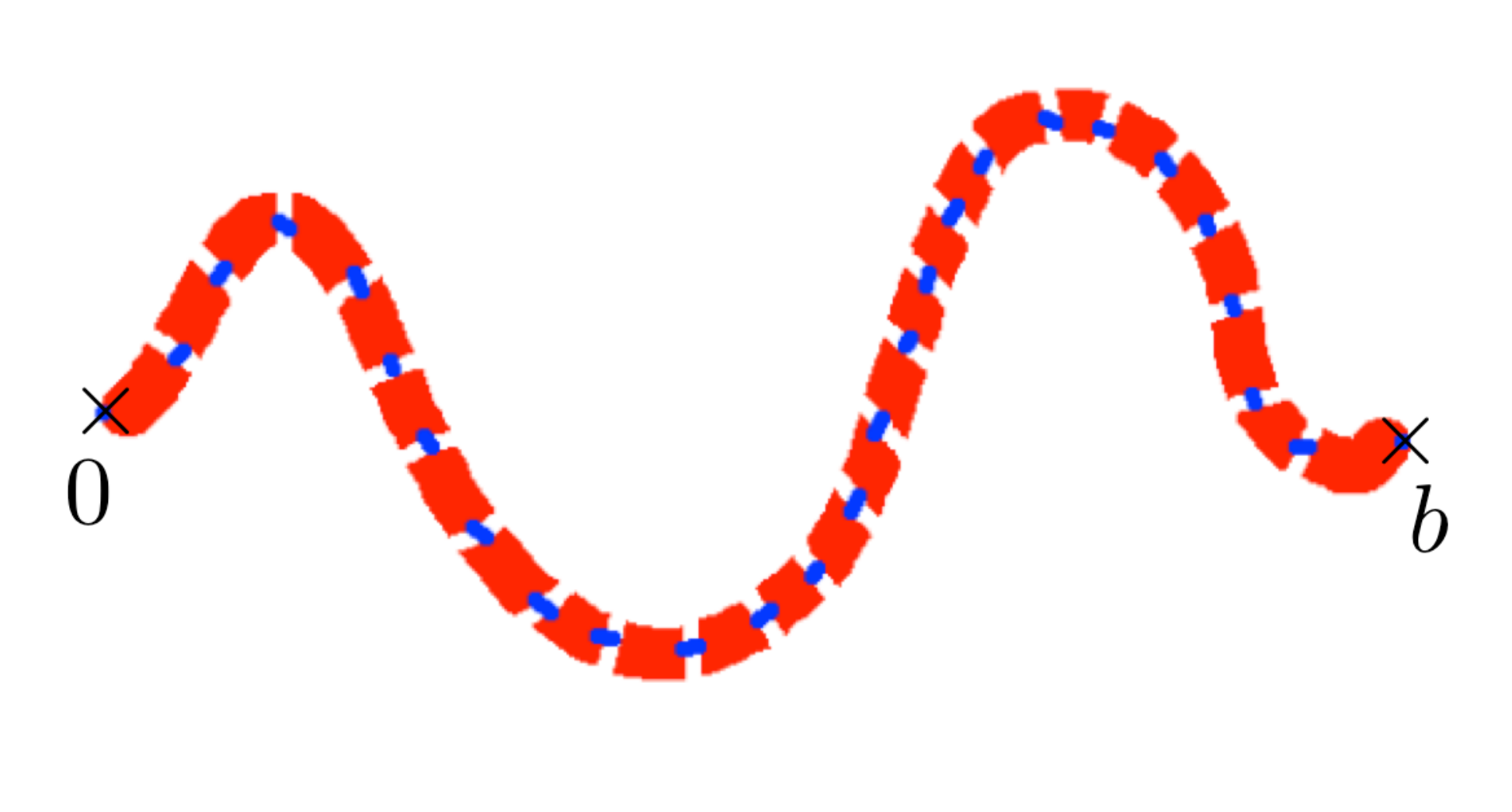}
    \includegraphics[width=6cm]{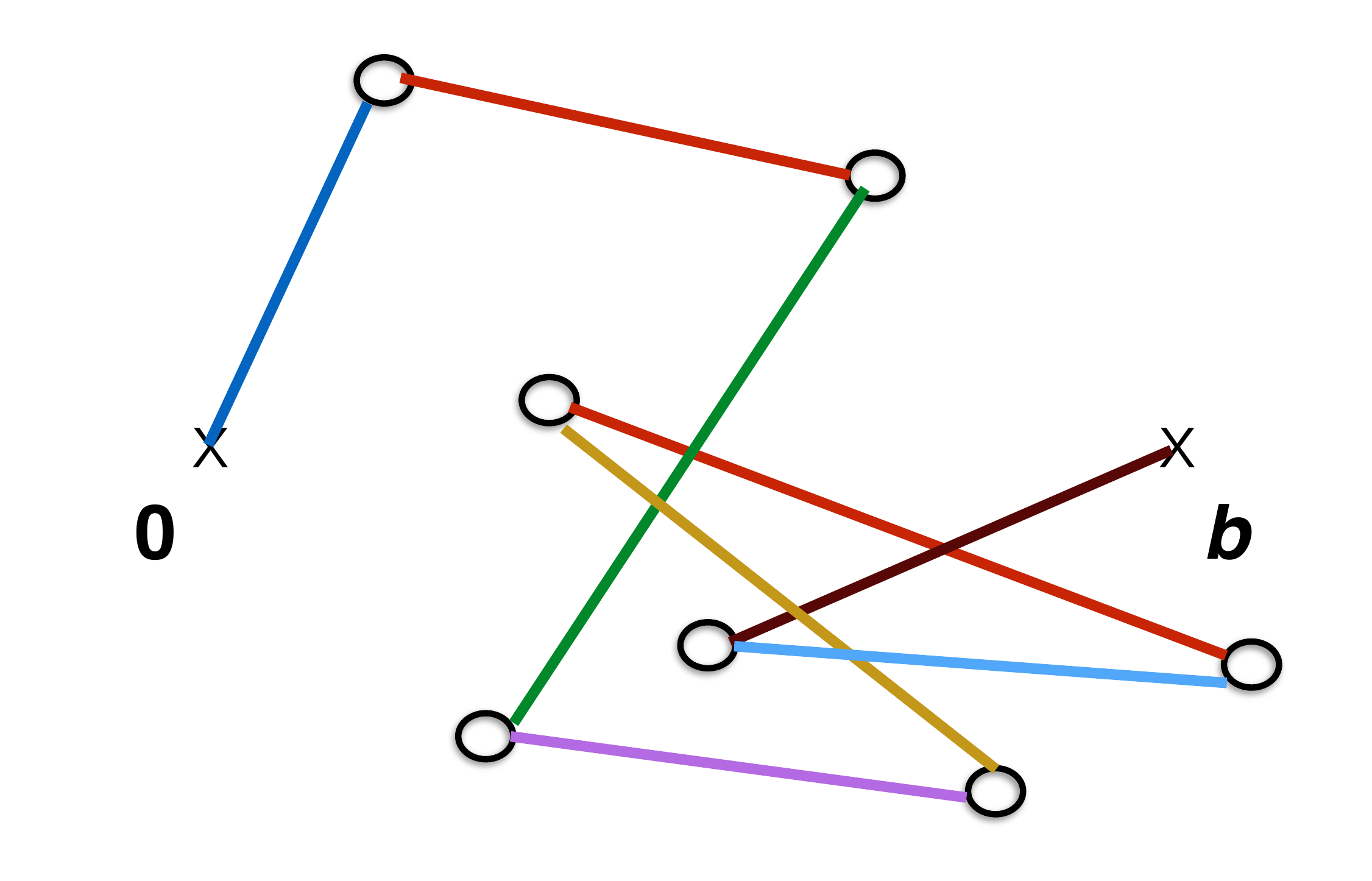}
  \caption{(a) String bits  in a stringy Pomeron,
  exchanged at fixed impact parameter $b$. (b) Gluons are indicated by small circles in the transverse plane, as emitted
  from a BFKL ladder diagram. In the limit of a large number of colors $N_c\rightarrow \infty$, the colors are unique
  for each gluon, and 
  there is one unique way to connect them by color fluxes. Since in pQCD there is
  little correlation between the gluon positions, the overall string length needed is much longer.
   }
    \label{fig_beed}
  \end{center}
\end{figure}

This density describes a cloud of string bits, in analogy to the cloud of gluons. However,
there is a significant difference: the string bits
  are not independent,  in fact they  together form 
 continuous QCD strings, stretched between the 
probe and the target dipoles, as shown in the lower plot  of Fig.~\ref{fig_tube} and also
in Fig.~\ref{fig_beed}a.

In the perturbative BFKL ladder diagrams, on the other hand, the produced gluons are
ordered only in their longitudinal rapidities. While their transverse locations 
 are also defined relative to those of the neighboring ``rungs" of the ladder, the corresponding
 integrals have a logarithmic measure $\sim dx_\perp^2/x_\perp^2$
allowing large jumps,  illustrated in Fig.~\ref{fig_beed}b. 
Instead of a continuous string, pQCD predicts essentially randomly placed 
gluons, with  color indices uncorrelated with positions. This create practical
problems and ambiguities in event generators, such as the Lund model
and its descendants.  

The density of string bits can be found in closed form
for AdS without and with a wall as we detail in appendix C. For small dipoles sizes or
${{\bf b}_\perp^2}/{2zz'}\gg 1$, 
it takes the simple form

\begin{eqnarray}
&&{\bf N}(\chi, z,z', {b}) \approx  2 \frac{e^{\Delta(\lambda) \chi}}{\left(4\pi \overline{\bf D} \chi\right)^{3/2}}
\nonumber\\
&&\times \frac{zz_0^2 }{z'{b}^2} {\rm ln}\left(\frac{{b}^2}{z z'}\right) 
e^{-{\rm ln}^2\left(\frac{{b}^2}{z z'}\right)/(4\overline{\bf D} \chi)}  
\label{WEEX}
\end{eqnarray}
with 

\be
\overline{\bf D}=\frac{\bf D}{z^2_0}\rightarrow \frac 1{2\sqrt{\lambda}}
\ee
${\bf N}$ is  the density of {\it string bits} of size $z$ at a transverse
distance $b=|{\bf b}_\perp|$ sourced by a dipole of size $z^\prime$ located at the origin. 
In Fig.~\ref{densitycomp} we show the density distribution  of  string bits (\ref{WEEX}) as the middle-blue-dashed  curve
as a function of the distance $b/z$
in the transverse plane, for $\chi=10$, $z=z^\prime=0.9\,z_0$  and   $\lambda=23$. The lower-orange-full curve 
is the unexpanded distribution without the AdS wall or the first contribution in (\ref{NEU}), and the upper-green-long-dashed curve is the full distribution in walled AdS 
as given in (\ref{NEU}).  The total number of  string bits   is

\be
{\bf N}=\int\,\frac {dz}{z}\,\frac{d{\bf b}_\perp}{z_0^3} \,\frac{z^\prime\bf N}{z_0} =e^{\Delta(\lambda)\chi} \, 
\label{HW4}
\ee
The total number of string bits thus grows exponentially with $\chi$, or as a power $\Delta$
of the collision energy.
Note that the number and distribution of string bits  (\ref{WEEX})  is conspicuously similar  to the distribution of  small dipoles in
the perturbative BFKL equations~\cite{Mueller:1994jq}
(again, modulo the substitution of the   Pomeron intercept).

While it is not really necessary, let us at the end of this section
present another --
perhaps  more intuitive --  way to visualize 
 this virtual cloud of string bits in  transverse space.
For a string with fixed end-points the transverse string coordinate $x_\perp^i(\sigma, \tau)$ can be 
discretized into $N+1$ points located at $0\leq \sigma=k\pi/N\leq \pi$ with $k=0, ..., N$ for each $i=1,..., D_\perp$,

\be
x_\perp^i \left(\frac {k\pi}N, \tau\right) = \frac{\bold{b}^i}{\pi}\frac{k\pi}{N }  +   \sum_{n=1}^{N-1}  X_{n}^i (\tau) \sin \left( n\frac{k \pi}{N } \right)  
\label{N1}
\ee
The string amounts to a collection of $N$ string bits, 
 or beeds  in the transverse space 
as illustrated in Fig.~\ref{fig_beed}. A similar interpretation for open strings on 
the light-cone  was suggested in~\cite{Karliner:1988hd}.
Normalization of the number of string bits  is the same as counting the number of stationary waves.
(Note that this counting is reminiscent of Debye counting of phonons
in solids.)

For large $b$, the Hamiltonian for the amplitudes $X_n^i$ 
of the stationary modes, follows  from the Polyakov-Luscher action as a collection of $N-1$ free oscillators in $D_\perp$ dimensions

\be
  \frac{1}{2 } \sum_{n=1}^{N-1}    \left(       \dot{X}_{n}^i (\tau)\dot{X}_{n}^i (\tau)   + 
n^2{X}_{n}^i (\tau)  {X}_{n }^i (\tau) \right)  + \frac{  b^2}{  \pi^2 } 
\label{HAR}
\ee
The ground-state  wave-function for this dangling string  is made of zero point contributions

\be
\label{ZERO}
\Psi_N[X]=\prod^{D_\perp}_{i=1}\prod^{N-1}_{n=1} \left( \frac{n^2}{\pi}  \right)^{\frac{1}{4}} 
\exp\left[  - \frac{n^2}{2} (X_n^i)^2   \right]
\ee
where all configurations with any number of string modes 
$n$ are equally probable.  Therefore the probability per configuration 
is $p_n=1/(N-1)$. The zero-point fluctuations on the string are maximally entangled, with an intrinsic
 von Neuman entropy

\be
{\cal S}=-\sum_{n=1}^{N-1}p_n\,{\rm ln}\,p_n= {\rm ln}\,{\bf N}_{wee}
\ee


Finally, let us briefly consider the process of $hadronization$. In a perturbative
approach this leads to a well known difficulty: gluons have color indices,  and 
eventually need to confront color confinement at large distances. Event generators as used in practice,
 such as the Lund model and its descendants, use certain stochastic algorithms for
 connecting gluon color indices by strings. The number of possible connections is very large and
 these algorithms are rather arbitrary. Furthermore,
 their modifications -- known as ``color reconnections" -- do affect the results.

In the stringy Pomeron model one can  quantify such rules. We propose to use for the produced strings
the density of ``wee-strings" (\ref{eqn_N}), or its confined form in walled AdS in (\ref{NEU}).

\subsection{Large multiplicity events}

As we already mentioned earlier, 
 in practice  in $pp$ collisions it is so far impossible to determine the value of the impact parameter $b$  for inelastic events. 
 Therefore,  one  needs to integrate over $b$,
 and, since our phase diagram suggests three different regimes for different
 $b$, the question is which one of the three, ``black", ``gray" or ``dilute", is the dominant one, for the high multiplicity tail.
 
 According to Kharzeev and Levin~\cite{Kharzeev:2017qzs}, it is the perturbative small-$b$ domain.
 According to Bjorken, Brodsky and Goldhaber~\cite{Bjorken:2013boa},
 it is the large-$b$ or cold string domain. (This is motivated by the fact that
 in this case the elliptic deformation parameter is as large as it can be, $\epsilon_2\sim 1$,
 thus maximizing the elliptic flow.)


In contrast, we propose that the tail of the multiplcity distribution 
may be dominated by the ``gray" or near-critical regime in our Pomeron phase diagram
(Fig.~\ref{fig_b-chi_plot}).  In this case the amplitude, in this so called Hagedorn regime, is



\be
\label{F0}
{\bf K}_T\left(\beta, b\right)\sim e^{-\beta {\bf F}_0}
\sim e^{-\sigma\beta b\left(1-\frac{\beta_H^2}{\beta^2}\right)^{\frac 12}}
\ee
with a vanishingly small  free energy ${\bf F}_0$
in the amplitude.  However, there are both a
 large energy (string mass)  
$M=-\partial \beta{\bf F}_0/\partial\beta$,
and a  large entropy, which conspire to cancel out. The string length is large
 $L/\sqrt{\alpha^\prime}=\beta^2\partial{\bf F}_0/\partial\beta$ 

\be
L\sim M\alpha^\prime\sim \frac{b\sqrt{2}}{4\pi}{\left(1-\frac{\beta_H}{\beta}\right)^{-\frac 12}}
\ee
and the partition function takes  the Hagedorn form

\be
{\bf K}_T\left(\beta, b\right)\sim e^{\beta_HL/\alpha^\prime}e^{-\beta M}
\ee
in which long strings are distributed {\it thermally}  as $e^{-\beta M}$. We 
speculate that events with initially long strings also 
carry large hadron multiplicities at the end of the collisions $N_L=M\sqrt{\alpha^\prime}\gg 1$
when they hadronize. In the inelastic collision the string bits come out
of the critical Pomeron, like beeds coming out of a shattered neckless.
Each string bit is a colorless closed string or glueball of typical mass $1/\sqrt{\alpha^\prime}\sim 1$ GeV, 
which can break  into several pions,  
 for $T=1/\beta$ close to 
$T_H\sim 160$ MeV as defined in (\ref{HAGEDORN}).
The normalized distribution for the
events with large multiplicity $N_L$ is therefore thermal 

\be
\label{PRO}
P(N_L)=\frac 1{\overline{N}_L}e^{-N_L/\overline{N}_L}
\ee
after fixing $\beta$ to reproduce the mean charged particle multiplicity which by our stringy estimates is

\be
\overline{N}_L=\frac 23\times 7\times {\bf N}_{\rm wee}= \frac {14}3\left(\frac{s}{s_0}\right)^{\Delta(\lambda)}
\ee
The fluctuations of the multiplicity distribution following from (\ref{PRO}) 
are given by the thermal cumulants

\be
C_q=\frac{\left<N_L^q\right>}{\left<N_L\right>^q}=q!
\ee
If so, one finds values for the first four moments, $C_q=2,6,24,120,...$ 
also close to the experimental ones given at the beginning of the section, inside the errors.
We interpret this agreement as an argument, that the final hadrons $may$ come from
the near-critical string-balls. Admittedly, at the time of this writing there is no proposal how to separate 
our and Kharzeev-Levin scenarios experimentally.

\section{Summary and discussion}

\subsection{Summary}
Before we summarize the paper, let us once again 
remind the reader of what was known before it.

The most nontrivial fact is that the  scattering amplitude has the same
Pomeron  form, both in the weak coupling BFKL theory, and in the stringy BKYZ one.
Gribov diffusion of gluons turns out to be
 quite similar to a diffusive propagation of a virtual string.
Both makes the hadron sizes grow at high energies. 
This has been shown in~\cite{Basar:2012jb}.

The next important point, from \cite{Shuryak:2013sra}, is about the difference between
the two. The stringy tube has a periodic coordinate, and thus an effective temperature and entropy. Furthermore, one can identify 
the existence of the new intermediate regime, related to Hagedorn-like string excitation.
It has been identified in the elastic profile $F(b)$. 

Kharzeev and Levin \cite{Kharzeev:2017qzs} described  gluon production by a
kinetic equation, and pointed out that the gluons produced have a very wide distribution
in multiplicity, very similar to the distribution over hadrons. As the distribution is calculated, one can also calculate the
``intrinsic entropy" of the system.  

Our main point now is that since in inelastic collisions there is (so far) no means to define
the impact parameter $b$, one has to integrate over it. As a result, 
{\em all three} regimes of the BKYZ Pomeron, as a function of $b$, seen
in elastic collisions, should also contribute to its intrinsic entropy.
The distribution in the number of gluons in pQCD domain needs to be
complemented by a study of the distribution over the
lengths and shapes of the promptly produced ``string balls" at intermediate $b$
and the ``cold strings" at large $b$.

We find the intrinsic entropy of the string to be very similar to that found perturbatively.
We found that the distribution over string-bits multiplicity is also very wide, and its  moments are also close
to the experimental ones. So, whether the  observed high-multiplicity events  originate from the gluonic cascade or long near-critical string balls, remains to be studied further.

\subsection{How smooth is the transition between the weak and strong coupling regimes?}

 While above we have emphasized the similarities between the  BFKL and BKYZ 
Pomerons above, now is the time to discuss their differences. While the entropy 
may be similar, only the second one has an effective temperature.
Furthermore, only the second one hints at the existence of the intermediate near-critical
regime, and on the fact that two sides of our ``phase diagram" are separated by something
resembling a  first order transition, like in thermal gluodynamics.

 The question at the title of this subsection can be specified further: Do all the Pomeron parameters 
 join  smoothly,   or can there be  some observable
remnants  of the Hagedorn phase transition? 

In pQCD the Pomeron intercept  in the leading order is 
\be 
\label{HARD}
\Delta_{BFKL}=\frac{4}\pi \,\alpha_sN_c\, {\rm ln}\,2
\ee
The next order $\alpha_s^2 N_c^2 $ correction \cite{Fadin:1998py} gets substantial
at $\alpha_s\approx 0.08$ or 't Hooft coupling $\lambda=g^2 N_c \approx 3$. 

In the stringy BKYZ version  in the holographic setting  and large $\lambda$, the  intercept
is (\ref{INTERCEPT})
  \be 
  \label{SOFT}
  \Delta_{BKYZ}={D_\perp \over 12}-\frac{(D_\perp-1)^2}{8\lambda}
  \ee 
where the number of transverse dimensions is holographically $D_\perp=3$. The second contribution 
contains the large coupling $\lambda=4\pi\alpha_s N_c $, is a correction induced by the curvature of AdS. 
Note that (like all one-loop semiclassical effects) the first leading term is independent
of the coupling and is just a number $\Delta=\frac 14$, being close to the empirical power
observed in $pp$ and DIS. 
 Note further, that this limit is reached at very strong coupling {\em from below}. So, this  
trend agrees at least in sign with the weak coupling BFKL result, and in principle those two can join smoothly.  
%
A rough estimate of the
cross-over coupling $\alpha_{s\star}$ for which this is expected to take place can be inferred 
from the leading order contributions if there is no jump in the intercept, 
\be
\alpha_{s\star}\sim \frac{\pi D_\perp}{48N_c{\rm ln}2}\sim 0.1
\ee

\begin{figure}[!htb]
\minipage{0.48\textwidth}
\includegraphics[width=60mm]{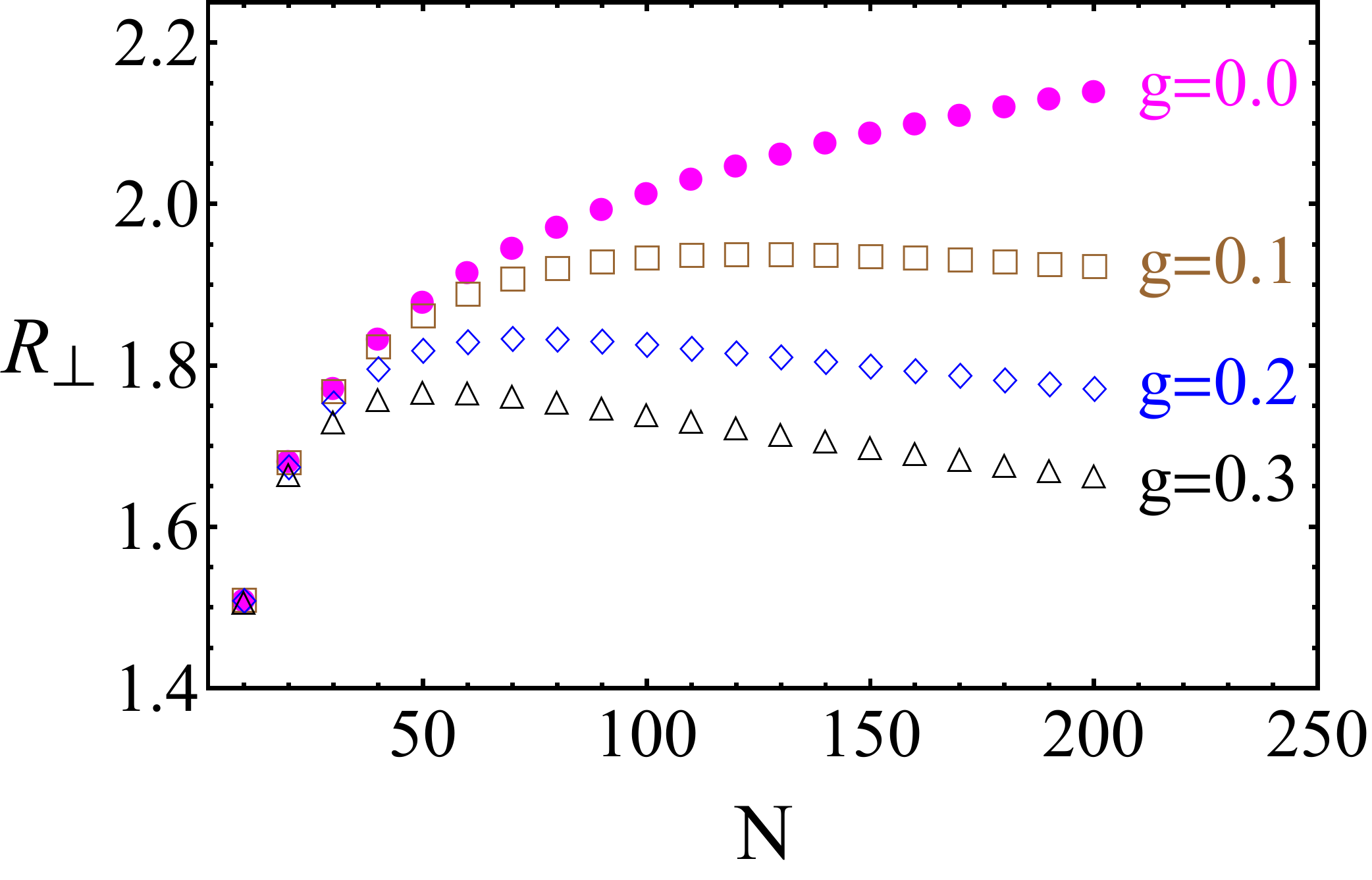}
\endminipage\hfill
\minipage{0.48\textwidth}
\includegraphics[width=60mm]{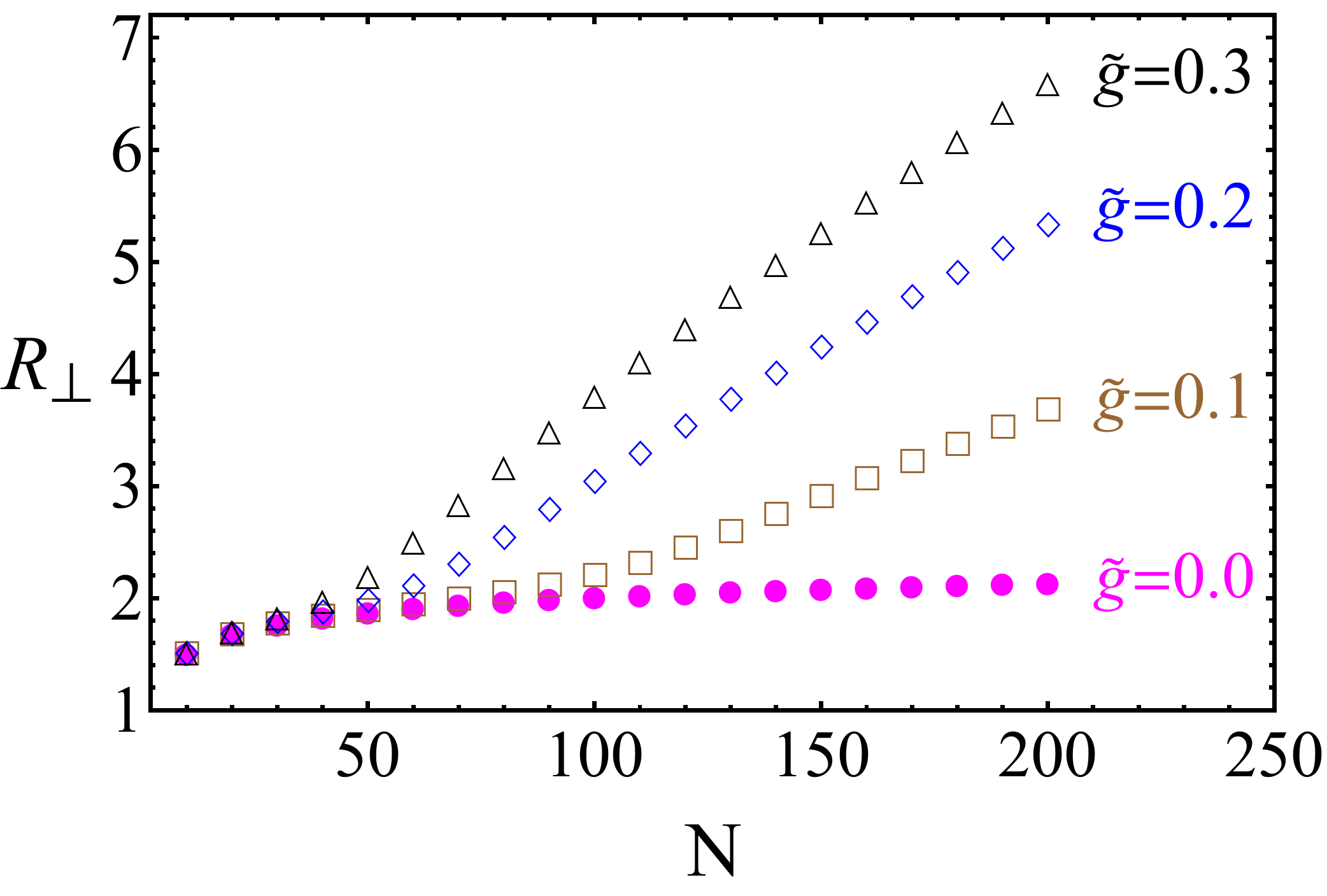}
\endminipage
 \caption{String transverse sizes for attracting self-interactions (top) and repulsive
 self-interactions (bottom) from~\cite{Qian:2015boa}.}
 \label{radius}
 \end{figure}

  \subsection{The  size and shape of the Pomeron and the interacting string balls}

A very massive  string can only behave as a black hole
if its self-attraction is taken into account
\cite{Susskind:1993ws,Damour:1999aw}. The objects, interpolating between free strings at small masses
and black holes at large ones, are known as {\em string balls}. 
Their generic properties in the string theory context have been given by
Damour and Veneziano~\cite{Damour:1999aw}. 

Its application to 
high multiplicity hadron collisions has been proposed in~\cite{Shuryak:2013sra}.
Detail studies of QCD string balls started in~\cite{Kalaydzhyan:2014tfa}. In it the evidences for self-interaction
of QCD strings from lattice studied have been reviewed, with the conclusion that it is, like gravity,
attractive, and is generated by an exchange of (scalar isoscalar) $\sigma$-meson exchange.
Unlike gravity, it has a finite range $1/m_\sigma \approx 1/600 \, {\rm MeV} \sim 0.3 \, {\rm fm}$.
The same conclusion has been reached in~\cite{Iatrakis:2015rga}, in which  the
holographic  AdS/QCD model  has been used to describe the QCD strings and their interactions. 
See also a holographic model discussed in~\cite{Liu:2014qrt}.

Numerical simulations of the QCD string-ball properties have been performed in 
\cite{Kalaydzhyan:2014tfa,Qian:2015boa}, via certain discretized models. 
One can reproduce strong free string excitations near the Hagedorn
temperature $T\rightarrow T_H$, and then probe how self-interactions affect
the string system. Another issue studied there has been shape fluctuations
of the string balls, described by azimuthal angular moments $\epsilon_m=\left<{\rm cos}(m\phi)\right>$,
which (for large enough multiplcity) can be converted by collective flows to observed 
angular moments of particle distributions $v_m$.


Here we would only describe, in 
  Fig.~\ref{radius},   how
 the transverse sizes of a self-interacting stringy Pomeron depend on the interaction
 strength, from the
exploratory study in~\cite{Qian:2015boa}. It  is based on the discretized form (\ref{N1}-\ref{ZERO}).
When the string self-attracts, the (squared) transverse  size 
changes from being proportional to the effective time ($\chi\sim {\rm log}(s/s_0)$)
 growth, typical of diffusion,  to a  fixed size. It  eventually shrinks into a collapsing regime, if 
the attraction is too strong. One can therefore speculate, that at some superhigh energy,
Gribov diffusion  and the expansion of the Pomeron size may stop.

Although we do not think this is the case, let us also explore the opposite case of a repulsive string-string
interaction.
The size of the self-repelling string was found  in~\cite{Qian:2015boa} to grow linearly with $N$. If so, the growth of the Pomeron size will continue even more rapidly than 
via Gribov diffusion.

 While there are observed showers from cosmic ray interactions 
in the atmosphere at $\sqrt{s}$ exceeding LHC by few orders of magnitude, 
low statistics and uncertainty about the composition of these cosmic rays
does not so far allow us to quantify the corresponding $pp$ cross section.

The final subject we would like to discuss is the intrinsic entropy and distribution of the string bits.  
In the critical regime of self-attracting strings, 
when the size levels off with increasing $N$, the transverse string density is
\be
\label{DENS1}
n_\perp=\frac {N}{R^{D_\perp}}\rightarrow \frac 1{g_{s}^2}\sim \frac 1{ l_P^{D_\perp}}
\ee
where the rightmost result follows at the critical value of $g_{s}^2N\sim 1$ with $R_\perp\sim N^0$. 
If we recall that in holography the Planck length follows from $l_P^{D_\perp}\sim g_s^2\sim G_N$ in $1+D_\perp$
spatial dimensions, (\ref{DENS1}) corresponds to one wee-string per Planck volume. A critical self-attracting 
string saturates the Bekenstein bound for the entropy ${\cal S}_B\sim N$ per unit area $A_\perp\sim R^{D_\perp}$,  with one bit per Planck volume.
The maximally entangled self-attracting string is a black hole~\cite{Susskind:1993ws}. 
At saturation, $g_s^2\sim l_P^{D_\perp}\sim1/N$ and the density of wee
strings reads

\be
\label{DENS2}
n_\perp\sim N\sim {{\bf N}_{\rm wee}}=\left(\frac 1x\right)^{\Delta(\lambda)}
\ee
At saturation, the proton size no longer grows with 
rapidity $\chi={\rm ln}(\frac 1x)$. It is a black disc with fixed edges. We note that in the opposite case of a self-repelling string, the 
transverse size  increases linearly with $N$ in strong violation of the Froissart bound.

For what critical  $x_\star$ one may expect to reach the black-hole limit  with
the cross section levelling off? While this value depends critically on 
the choice of $g_s$  and the details of the string self-interaction,
for the weakest coupling $g_s\sim 0.1$, the  exploratory studies 
shown in Fig.~\ref{radius}  (flat space), suggest a transition 
for $N_\star\sim 100$. Using (\ref{DENS2}) this translates to
$x_\star$ 
\be
\label{XC}
x_\star\sim \left(\frac 1{N_\star}\right)^{\frac 1{\Delta}}\sim 10^{-8}
\ee
The relative rapidities currently reached at the LHC with $pp$ collisions at $\sqrt{s}\sim 7$ TeV and for $\sqrt{s_0}\sim 1$ GeV,
correspond to a relative rapidity of $\chi={\rm ln}(s/s_0)\sim 18$.  
This translates to a cloud of string bits  at a resolution of
$x\sim 10^{-8}$.

  \vskip 1cm
{\bf Acknowledgements.} We  thank  Dima Kharzeev  for discussions.
This work was supported by the U.S. Department of Energy under Contract No. DE-FG-88ER40388.
\vskip 0.5cm

    \appendix
 \section{Static potential and duality of string quantization}
The potential energy of a static quark-antiquark pair, separated by a distance $r$ has been studied since the birth of QCD, 
 via lattice simulations and quarkonia phenomenology. It is well known that
  at small distances it has the perturbative Coulomb-like behaviour, with the running coupling
  $\alpha_s(r)$, while 
 at large distances it exhibits a linear confining potential $V_{\rm conf}=\sigma_T r$ due to a confining string. One question to be discussed here is where and how those two regimes meet.
 Another, more technical one, is whether the known expressions for the static potential
 can be related to our expressions for the dipole-dipole scattering amplitude.
 
 Specifically, one can address these questions by studying corrections to the
 leading term for the potential, both from small and large distances. We will follow the latter and
recall that the quantum vibrations of the QCD string leads to 
   the so called ``Luscher term" \cite{Luscher} 
 \be V_{\rm Luscher }(r)=\sigma_T r-{\pi \over 12}{1\over r}+...\ee  with
 a new Coulomb-like  correction, to be valid at large $r$.
 Re-summation of the quantum corrections of the Nambu-Goto string leads to the so called Arvis potential ~\cite{Arvis:1983fp}
 \be    V_{\rm Arvis }(r)=\sigma_T r \left(1-{\pi \over 6} {1 \over \sigma_T r^2}\right)^{\frac 12}
 \ee
This form suggests a singularity at a distance at which the bracket vanishes.
Clearly, the string description is invalid beyond this point, as it cannot produce a negative energy.  
The static potential of course does change sign, becoming negative at $r< 0.2\, {\rm fm}$. 

Its behavior   in the intermediate region has been studied on the lattice, at zero and non-zero temperatures, see e.g. \cite{Petreczky:2005bd}. One can use Fig.~6 of this work to answer
the following two questions: (i) at which distances the effective $\alpha_s(r)$ gets smaller than
the Luscher value $(3/4)(\pi/12)\approx 0.2$?; and (ii) at $T>T_c$, when  the confining string is absent, at which distances one finds the same 
 $\alpha_s(r)$? The answer to both is the same, namely $r < 0.15 \,{\rm  fm}$.  One can use
 it as a definition of the applicability region of pQCD, for this problem.
 
 Before turning to more technical issues, let us note that  the static potential is not
 directly observable: spectra for quarkonia were fitted with a Cornell potential, in which
 $\alpha_s\sim 0.5$.  Distances $r \sim 0.15 \, {\rm fm}$ do not play a particularly important role in 
 the spectroscopy. The scattering amplitude profile $F(b)$, on the contrary, is
 directly observable, and it basically contains the potential energy in the exponent. That is why one can
 detect in it a pQCD-string transition much better than in the static potential.
 
        \begin{figure}[t]
  \begin{center}
  \includegraphics[width=8cm]{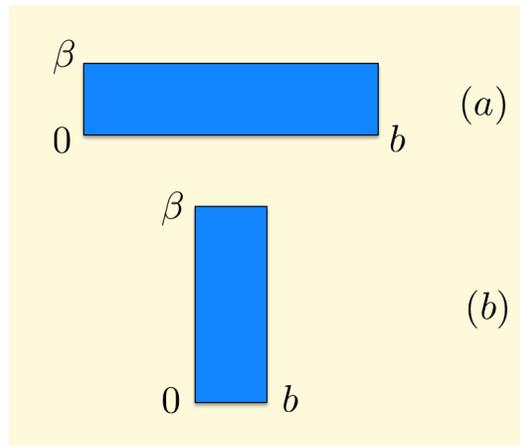}
  \caption{The world-sheet of a Pomeron  as a close string  moving along the long  b-direction (a), and a
   static quark-antiquark pair linked by  an open string  moving along the long $\beta$-direction (b).
  The world-sheets  shown as  rubber strips are periodic in $\beta$.}
    \label{rubber}
  \end{center}
\end{figure}

 Now we turn to the
   relationship between the expressions for the Pomeron and the static potential.  
%
 In Fig.~\ref{rubber}a we sketch the world-volume of the exchanged string in the Pomeron.
 Its spatial extent is the impact parameter $b$, and its Euclidean time extent is $\beta$
 (\ref{eqn_beta}). The semiclassical derivation assumes ``cold string", with $b>\beta$.
   In Fig.~\ref{rubber}b we sketch the 
 open string between a heavy quark-anti-quark pair propagating along the long  and periodic 
 $\beta$-direction. In this case it is assumed that the temperature goes to zero, or $b\gg \beta$.
 
 The interplay between the Pomeron and the potential 
 expression  reflects  the
 general conformal nature of the string action under the exchange of
 Euclidean time and space coordinates $\tau\leftrightarrow \sigma$.
 In doing so, however, one needs to pay attention to different
  boundary conditions along time and space. The string is always expected
  to be $periodic$ in time, but in space it has rigid boundary conditions,
  as the string is attached to locations of the Wilson lines.

The Nambu-Goto evaluation \cite{Arvis:1983fp}, corresponding to Fig.~\ref{rubber}b,  yields a
partition function

\begin{eqnarray}
\label{14X00}
&&{\bf Z}_T\left(\beta, b\right)=\nonumber\\
&&{\rm Tr}\left(e^{-\sigma\beta b\left(1+\frac{24}{D_\perp}\frac{{\beta}_H^2}{ 4b^2}\left({\bf L}_0-\frac{D_\perp}{24}\right)\right)^{\frac 12}}\right)\qquad{}
\end{eqnarray}
It is readily seen that  (\ref{14X00})  follows from  (\ref{14X0}) and vice se versa, under the exchange $$2b\leftrightarrow \beta$$ 
The meaning of the factor 2 here is precisely due to the fact that the rigid boundary condition
corresponds to ``half-circle", and only becomes a full circle (or rather a tube) if complemented by another ``half-circle". In more precise words, in order to have exact space-time duality
of the two problems,
one needs to double the spatial extent to have a topology of the double-torus. 

 Both string descriptions require the strings not to be highly excited, or $\beta>\beta_H$ 
 as illustrated in the Pomeron phase diagram in Fig.~\ref{fig_b-chi_plot}. (\ref{14X00}) holds for Fig.~\ref{rubber}b
 for $\beta>b$ or $\chi<2\pi$ (small rapidity), while (\ref{14X0}) holds  for Fig.~\ref{rubber}a for 
 $\beta<b$ or $\chi>2\pi$ (large rapidity).

\section{The regime dominated by ``Pomeron daughters"}
In the phase diagram, for simplicity, we have only shown one region, dominated by
the near-critical string balls. However, in the lower left corner of it there exists a separate
regime dominated by a different type of excitations, related to the so called  ``Pomeron daughters". The amplitude
(\ref{14X0})  includes 
a sum not only over the zero point oscillations of the
string, but also over its thermal excitations, of the type

\be 
{\bf K}_T(\beta, b)\sim 
\sum_n d(n) e^{- \sigma \beta b \left(1- {\beta_H^2 \over \beta^2}+ \frac{8\pi n}{\sigma\beta^2}\right)^{\frac 12} } 
\label{eqn.K}  
\ee   
where $d(n)$ is the density of excited states. In so far we assumed $n=0$ and ignored
the Pomeron daughters with $n=1,2...$. Now, consider what happens if $\beta < \beta_H$
when the Pomeron daughters dominate. In this case too, the strings carry large masses
 $M_n\sim \sqrt{n/\alpha^\prime}$ so that (\ref{eqn.K}) becomes

\be 
{\bf K}_T\sim \sum_n d(n) e^{- 2 b \sqrt{2\pi\sigma n }} \sim \sum_n e^{(\beta_H-2b)M_n }
\ee
where we used  that $d(n)\sim e^{\beta_HM_n}$. So, as the distance $2b$  approaches $\beta_H$
there is a separate Hagedorn transition, now induced by the multiple excitations of the Pomeron daughters. 

This effect is, however, only important for low collision energies and small impact parameter $b < \beta_H/2$. Otherwise, as we discussed in the main text, 
the tachyon-induced second term in (\ref{eqn.K}) is dominant.
  In all cases, the long
string distributions are thermal  with an intrinsic  
temperature of either $1/\beta$ or $1/2b$, but the same Hagedorn temperature (\ref{HAGEDORN}).\\

\section{Diffusion in modified $AdS_5$  }
In the AdS/QCD models 
the holographic string exchange takes place in  $AdS_5$  space,
 modified by a certain profile of the dilaton field. The simplest version
describing confinement, makes use of a ``confining wall", by cutting off
 space beyond a certain value of the holographic coordinate.
The quantum version of (\ref{14X1}) in a hyperbolic slice of AdS$_{D_\perp}$ 
with line element

\be
ds_\perp^2=\frac{z_0^2}{z^2}(d{\bf b}_\perp^2+dz^2) \ ,
\ee
confined to  $0\leq z\leq z_0$,  follows  in the form~\cite{Stoffers:2012zw}

\be
\left(\partial_{\chi}-\Delta-\frac {\bf D}{\sqrt{g_\perp}}\partial_\mu\,g_\perp^{\mu\nu}\sqrt{g_\perp}\,\partial_\nu\right)\,
{\bf K}_T (x_\perp, x'_\perp)=0 \ , 
\label{TACH3}
\ee
with the short notation $x_\perp=(z, {\bf b}_\perp; \chi$). 
Using the conformal variable $u=-{\rm ln}(z/z_0)$,  the initial condition of one wee string
at ${\bf b}_\perp={\bf b}^\prime_\perp=0$, and the zero current condition 
$\partial_z{\bf N}=0$ at the wall  boundary $z=z_0$ (no leak at the wall), the solution to (\ref{TACH3})
for the dimensionless density of wee strings ${\bf N}\approx {\bf K}_T/zz^\prime/z_0^2$ is

\begin{eqnarray}
\label{NEU}
&&{\bf N}(\chi,u,u^\prime,b)=\nonumber\\
&&e^{u^\prime+u}\,{\mathbb  K}_T(\chi,\xi)
+e^{u^\prime-u}\,{\mathbb  K}_T(\chi,\xi_*)
\end{eqnarray}
with

\be
{\mathbb K}_T(\chi,\xi)=\frac{e^{\Delta(\lambda)\chi}}{(4 \pi \overline{{\bf D}} \chi)^{3/2}} \frac{\xi 
e^{-\frac{\xi^2}{4\overline{{\bf D}} \chi}}}{\sinh(\xi)} \ ,
\label{PRO5}
\ee
The chordal distances  are

\begin{eqnarray}
{\rm cosh}\xi&=&{\rm cosh}(u'-u)+ \frac{{\bf b}_\perp^2}{2z_0^2}\,e^{u'+u}\, \\
{\rm cosh}\xi_*&=&{\rm cosh}(u'+u)+ \frac{{\bf b}_\perp^2}{2z_0^2} e^{u'-u}\,\nonumber \ ,
\label{HW6}
\end{eqnarray}
with $-u$ the image of $u$ with respect to the holographic wall at $u=0$ ($z=z_0$).  
The Pomeron intercept is now (\ref{INTERCEPT}).  The density of wee-strings for AdS
without a wall corresponds to only the first contribution in (\ref{NEU}).

\end{document}